\documentclass[fleqn,10pt]{wlscirep}
\usepackage[utf8]{inputenc}
\usepackage[T1]{fontenc}
\usepackage{parskip}
\usepackage[show]{chato-notes}% adri
\usepackage{comment}
\usepackage{subcaption}

\title{Network and Sequence-Based Prediction of Protein-Protein Interactions}
\author[1,*]{Leonardo Martini}
\author[1,*]{Adriano Fazzone}
\author[1,*]{Luca Becchetti}

\affil[1]{Sapienza University of Rome, Italy}

\affil[*]{lastname@diag.uniroma1.it}

\begin{abstract}
\textbf{Background:} Typically, proteins perform key biological functions by interacting with each other. As a consequence, predicting which protein pairs interact is a fundamental problem. Experimental methods are slow, expensive, and may be error prone. Many computational methods have been proposed to identify candidate interacting pairs. When accurate, they can serve as an inexpensive, preliminary filtering stage, to be followed by downstream experimental validation. Among such methods, sequence-based ones are very promising.
\textbf{Results:} We present, a new algorithm that leverages both topological and biological information to predict protein-protein interactions. We comprehensively compare our Framework with state-of-the-art approaches on reliable PPIs datasets, showing that they have competitive or higher accuracy on biologically validated test sets.
\textbf{Conclusion:} We shown that topological plus sequence-based computational methods can effectively predict the entire human interactome compared with methods that leverage only one source of biological information.
\end{abstract}
\begin{document}

\flushbottom
\maketitle

\section{Introduction}
\label{sec:intro}

% Problem Definition and Goal of this Pre-Print 
Protein-Protein Interactions (PPIs) play a crucial role in several biological processes since, in many cases, proteins perform vital functions by interacting with each other in the formation of protein complexes. The identification of new Protein-Protein interactions is thus crucial in understanding cells' biological mechanisms. Furthermore, knowledge of the interactions can be exploited for applications such as drug repurposing \cite{}, which leverages network topology to predict drug-disease associations, or network-based approaches to disease-gene prioritization \cite{}, which leverage the PPI network to find new candidate disease genes. Consequently, charting protein-protein interaction maps remains a fundamental goal in biological research.

% Why developing computational algorithm is a crucial problem 
Protein-protein interactions can be most readily identified by protein affinity chromatography or pull-down experiments, yeast two-hybrid screens, or purifying protein complexes that have been tagged in \textit{vivo}. 
These methods are all labor and time consuming and have a high cost associated with them. Each of them has inherent advantages and disadvantages. For instance, the yeast two-hybrid system has the advantage of identifying the direct interaction between protein pairs. However, data gathered using this method may contain a high (as much as 50\%) rate of false positives. 
Therefore, in the absence of other lines of evidence, this data alone cannot be considered biologically significant. 
The high cost and the technical limitations associated with such biochemical approaches have resulted in a growing need for the development of computational tools capable of identifying prospective protein-protein interactions.

Several tools have been designed for this purpose over the past few years. Some of these approaches predict protein interactions using the primary structure of proteins themselves. In this line of research, SPRINT\cite{li2017sprint} and PIPE\cite{pitre2006pipe} are two well-known algorithms that rely on the same underlying hypothesis: a pair of proteins similar to a pair of interacting proteins have a higher chance to interact. In this line of thought, some very promising approaches, like L3\cite{} and SIM\cite{}, leverage network topology to define the notion as mentioned earlier of similarity.

% Introduction to our framework
We propose a new framework that can exploit topological and biological information to predict protein-protein interactions. The algorithm relies on the underlying hypothesis that two proteins interact in proportion to the structural similarity that one has with the most similar of the interactors of the other. From a topological perspective, the structural similarity between two proteins is proportional to the number of common neighbors. Instead, from a biological perspective, the structural similarity between two proteins is proportional to the similarity of their primary sequences.

% Validation our hypothesis 
We compare our framework with the state-of-art approaches on several synthetic and Human PPI networks. We show that it outperforms many heuristics in predicting protein interactions. Moreover, candidate protein pairs prioritized by our algorithm are involved in the same biological processes, molecular functions, and cellular components. Finally, a subset of the candidate protein pairs predicted by our framework has been experimentally validated by Xu-Wen Wang et al.
\section{Materials and Methods}
\label{sec:materials_and_methods}

\subsection{Scoring Protein Interaction}
\label{sub:scoring}
\begin{figure}[h]
     \centering
     \begin{subfigure}[t]{0.49\textwidth}
         \centering
         \includegraphics[width=\textwidth]{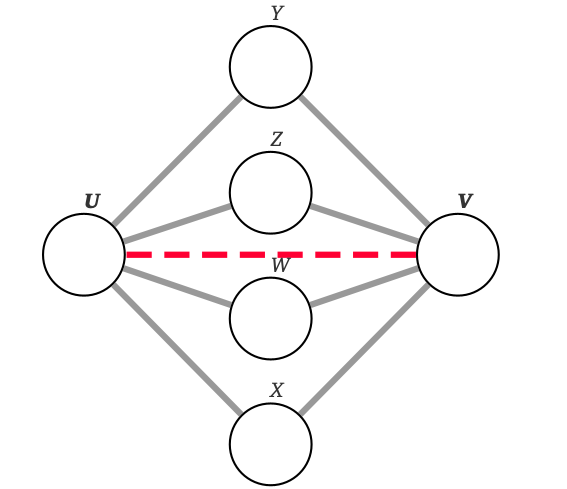}
         \caption{TCP predicts (P) links based on node similarity (S), quantifying the number of shared neighbors between each node pair $(A^2)$}
         \label{fig:JC}
     \end{subfigure}
     \hfill
     \begin{subfigure}[t]{0.49\textwidth}
         \centering
         \includegraphics[width=\textwidth]{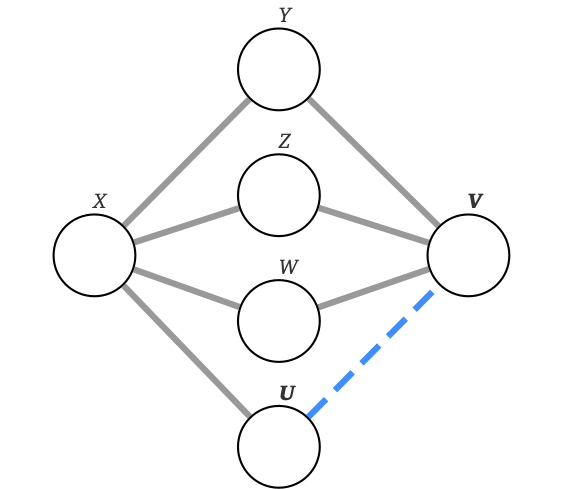}
         \caption{PPIs often require complementary interfaces. Hence, two proteins, X and Y, with similar interfaces share many of their neighbors. An additional interaction partner of X (protein U) might be also shared with protein Y (blue link)}
         \label{fig:L3}
     \end{subfigure}
        \caption{comparison of frameworks based on 2-path lengths and those based on 3-path length }
        \label{fig:JC_vs_L3}
\end{figure}

One way to detect protein-protein interaction consists of look at each interactor's tertiary structure to find complementary binding sites. However, compared to topological and protein's primary structure, the protein's shape information (tertiary and quaternary structure) is scarce. This lack of information is consequence of the expensive cost to discover the tertiary structure of proteins using new technologies. Network-based approaches, that score the likelihood to have a direct interaction between a protein pair, do not need any protein's structure information, but they can infer it leveraging network topology. Furthermore, the increasing coverage of the interactome has inspired the development of network-based algorithms, which exploit the patterns characterizing already mapped interactions to identify missing interactions. The problem of identifying new links in the interactome is known as link prediction problem \cite{}. To predict new links in the network several types of measures have been studies and proposed\cite{}. The most promising measures used in link prediction problem relies on two different hypothesis: the hypothesis that nodes topological similar to each others (i.e nodes that shares several neighbours ) and the hypothesis that a node should be linked a candidate if it is similar to the known interactors. Figure \ref{fig:JC_vs_L3} shows the different between measures that leverage the former hypothesis and those ones that exploit the latter idea.

More precisely, figure \ref{fig:JC_vs_L3} a) shows measures that scores a possible interaction between node $u$ and $v$ computing the number of neighbours $u$ and $v$ shares. Metrics that use this information are Jaccard Index, Common Neighbours, Adamich Adar, $CH1\_L2$ and $CH2\_L2$. While figure \ref{fig:JC_vs_L3} b) shows groups metrics that link $u$ and $v$ if $v$ is similar to $u$'s partner and vice-versa. Furthermore, metrics that leverage common neighbours have been widely used in link prediction problem on social network, while, metrics that leverage partner similarity are more biological driven and they are based on the hypothesis that if $v$ is similar to an $u$'s interactor, then $u$ and $v$ are likely to interact (i.e they shares complementary binding sites).

\begin{figure}[h]
     \centering
     \begin{subfigure}[t]{0.49\textwidth}
         \centering
         \includegraphics[width=\textwidth]{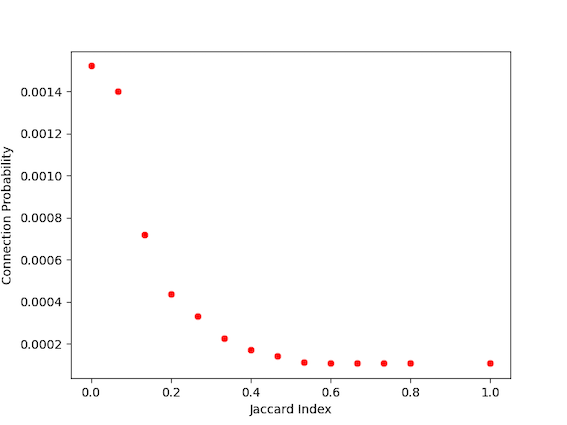}
         \caption{High Jaccard similarity indicates a lower chance for the proteins to interact }
         \label{fig:JC_connection_probability}
     \end{subfigure}
     \hfill
     \begin{subfigure}[t]{0.49\textwidth}
         \centering
         \includegraphics[width=\textwidth]{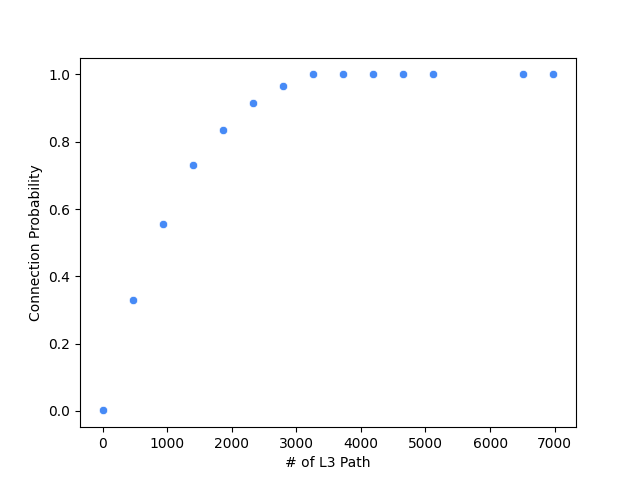}
         \caption{We observe a strong positive trend in HuRI between the probability of two interacting proteins and the number of $l=3$ paths between them}
         \label{fig:L3_connection_probability}
     \end{subfigure}
        \caption{Network similarity does not imply connectivity. \textbf{a} and \textbf{b} show the difference between TCP paradigm and L3 principle.}
        \label{fig:JC_vs_L3_connection_probability}
\end{figure}

To compare these two approaches that relies on different ideas, we computed the probability of a pair $(u,v)$ to be connected in the Protein-Protein Interaction (PPI) network when Jaccard Index (figure \ref{fig:JC_vs_L3_connection_probability} a)) and the number of three path length between $(u,v)$ (figure \ref{fig:JC_vs_L3_connection_probability} b)) are increasing. 
To estimate this probability, we first computed the Jaccard Index and the number of three path length of each pair of nodes in the PPI Network G(V,E). 
We estimated the probability of connection in the following way:
\begin{equation}
    P\left(\left(u,v \right) \in E \mid s \geq x\right) = \frac{\sum_{\left(u,v\right) \in V \times V} \chi_{u,v}}{\mid S \mid}
\end{equation}

Where $\chi_{u,v}$ is 1 if $(u,v) \in E$ and 0 otherwise, while S is the set that contains all pairs $u,v$ with a score $s \ge x$.

Figure \ref{fig:JC_vs_L3_connection_probability} demonstrate that, if we use as score the Jaccard Similarity fig. \ref{fig:JC_vs_L3_connection_probability} a) and the number of three path length fig.\ref{fig:JC_vs_L3_connection_probability} b), protein pairs with high Jaccard Similarity do not tend to interact if we compare them with $A^3$ (i.e. the number of 3 path length between $u$ and $v$). This pattern is visible on several Human and not Human PPI Network as shown in Supp. Fig. \ref{fig:connection_probability_all_network}

\subsubsection{Jaccard Index as a Measure of Protein Interface Similarity}
\label{sub:jac_as_protein_interface_sim}
As discussed so far, protein pairs with high jaccard similarity do not tend to interact. This could be due to the fact that proteins that shares several neighbours should have similar binding sites (i.e similar tertiary structure). To assert the validity of this hypothesis, we considered the correlation between sequence similarity and the Jaccard Index of a pair $(u,v)$. More formally, we define the Protein sequence Similarity of $(u,v)$ as the distance between the longest sequence and the global alignment of their protein primary structures computed using the Needleman–Wunsch algorithm \cite{}.

\begin{figure}[h]
     \centering
     \begin{subfigure}[b]{0.45\textwidth}
         \centering
         \includegraphics[width=\textwidth]{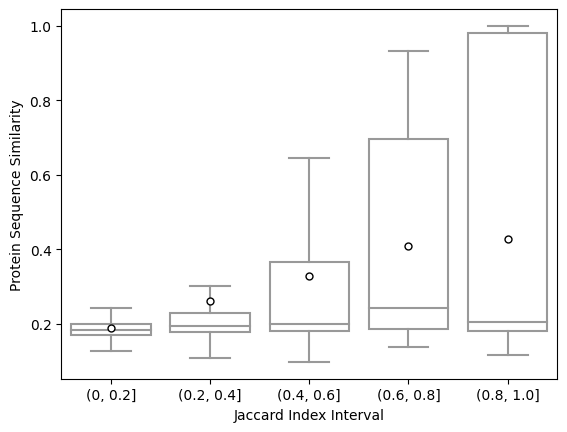}
         \caption{HuRI}
         \label{fig:HuRI}
     \end{subfigure}
     \hfill
     \begin{subfigure}[b]{0.45\textwidth}
         \centering
         \includegraphics[width=\textwidth]{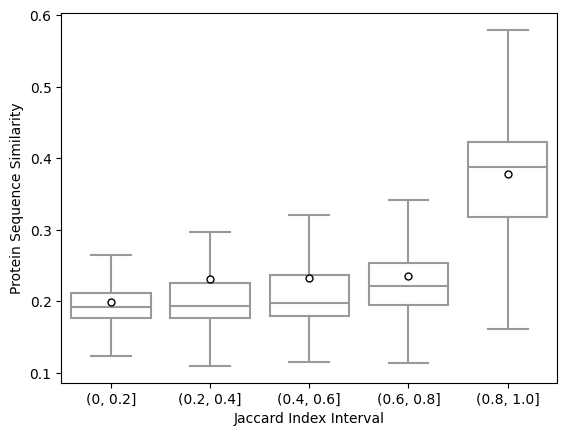}
         \caption{STRING}
         \label{fig:STRING}
     \end{subfigure}
        \caption{Jaccard similarity as a measure to infer protein structure similarity.}
        \label{fig:jaccard_index_sequence_sim}
\end{figure}

Figure \ref{fig:jaccard_index_sequence_sim} a) shows the correlation between Jaccard Similarity and Protein sequence similarity. To better understand the correlation between the score induced by the global alignment and the Jaccard Similarity of a protein pair, we grouped each pairs by their Jaccard Index creating four different buckets as shown in the X axes of fig. \ref{fig:jaccard_index_sequence_sim} a) . For each bucket, we drawn the box plot representing the overall distribution of the global alignment score of protein pairs with similar Jaccard Index. As illustrated in fig. \ref{fig:jaccard_index_sequence_sim} a), when the Jaccard Similarity is in the interval $\left(\frac{3}{4}, 1\right]$
, protein pairs shares similar sequences. 
Finally, fig \ref{fig:jaccard_index_sequence_sim} b) illustrates the number of samples belonging to each bucket. 
Every bucket, with the exception of the bucket with interval $\left(\frac{1}{2}, \frac{3}{4}\right]$, has a similar number of samples.

\subsubsection{Jaccard Index as a Measure of Gene Duplication Phenomena}
\label{sub:jac_as_gene_dup}
Jaccard Index, not only is useful to find proteins with similar structures, but it can be applied to identify proteins originated by  the process of  gene duplication. Indeed, in the process of evolution, genes may produce new proteins, which may retain many of the biological functions of the original ones. As we know, the structure of a proteins is related to its biological functions \cite{} and thus, proteins born from the process of gene duplication should share similar interactors. To statistically quantify if proteins created by this evolutionary process show this behaviour, we downloaded from \cite{ouedraogo2012duplicated} a data set consisting of groups of protein products generated by the gene duplication process. Firstly, we filtered out the smallest groups, keeping only groups consisting of  a large number of proteins (i.e number of proteins $\ge 10$ ) that appear in the PPI network. Secondly, for each group $I$, we defined the Mean Jaccard Index of group $I$ ($MJI_I$) as:
\begin{equation}
    MJI_I = \frac{1}{\mid m \mid} \cdot \sum_{(u,v) \in IxI} J(u,v)
    \label{eq:mean_jaccard}
\end{equation}

Where $m$ is the size of $IxI$. Finally, we compared the value of each group's Mean Jaccard Index with a random distribution of the score computed using random set of proteins with the same size of the original group $I$.

\begin{figure}[ht]
\centering
\includegraphics[width=\textwidth]{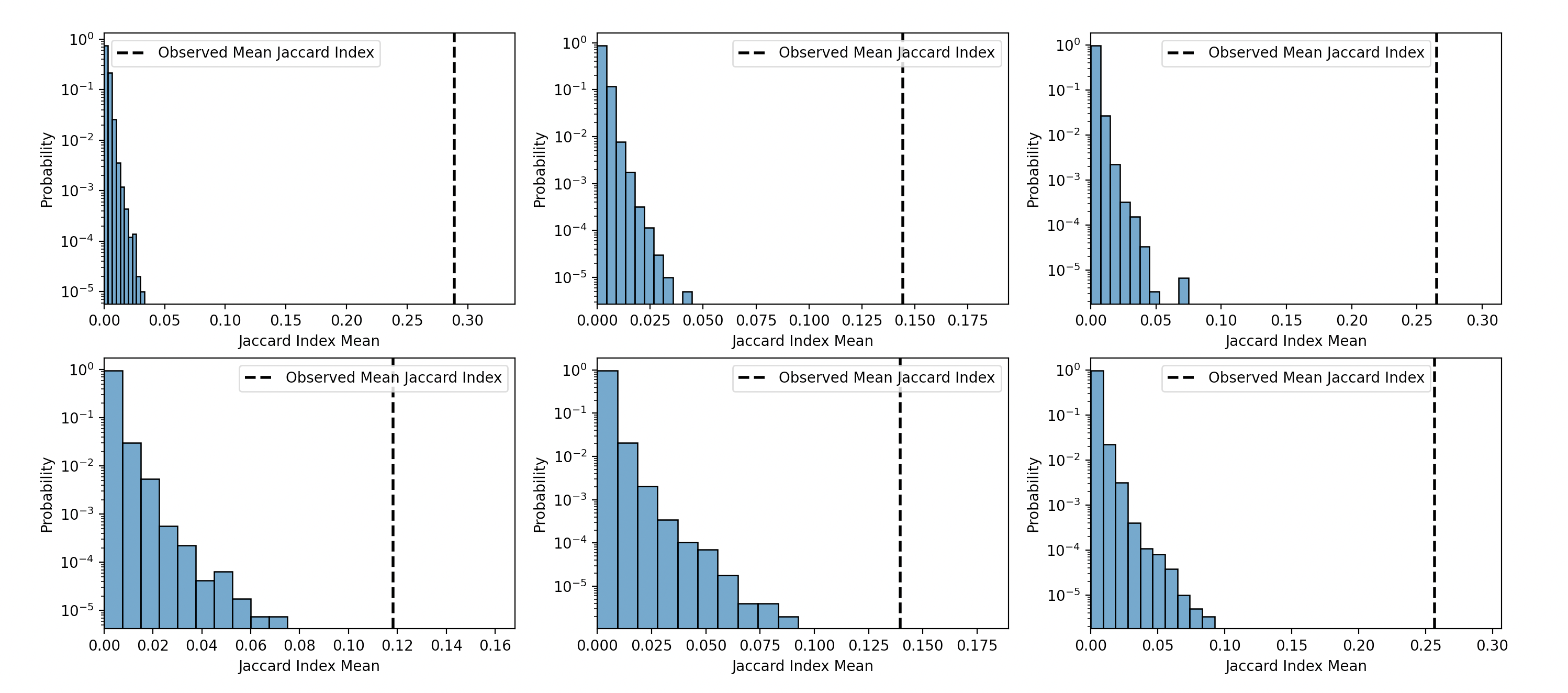}
\caption{Proteins originated by Gene Duplication Phenomena shares more common naighbors than Random Expectation.}
\label{fig:duplicated_gene_set}
\end{figure}

Figure \ref{fig:duplicated_gene_set} shows the scores of Mean Jaccard Index of the original groups and compare them with a random distribution. As illustrated by each plot, the value of the $MJI$ of each group is very large if we compare with the associated random distribution. Indeed, each vertical line, representing the score of each group is in the tail of each distribution. As, discussed before, each group consists of a number of proteins in the interval $[10, 27]$.

\subsubsection{Jaccard Index to Model Evolutionary and Functional Similarity}
\label{sub:jac_for_seq_sim}
The relationship between protein sequences and structures has long been a widely accepted tenet of biochemistry\cite{kidera1985relation,guzzo1965influence}. Indeed, one of the foundations of molecular biology is that a protein's sequence determines its structure, which determines how the protein functions. 

Protein sequence–structure-function relationships have been investigated and quantified in various ways. Several studies\cite{chothia1986relation,wilson2000assessing} have established the correlation between structural similarity and sequence similarity. Other ones\cite{rost1999twilight, yang2000integrated} have studied the level of sequence similarity at which structural similarity is likely to be observed. Consequently, protein primary sequence similarity indices have been widely used to capture evolutionary relationships between paralogs (i.e., homologous proteins related to a gene duplication phenomena)\cite{jeffryes2018rapid}. 

Consequently, we modeled a protein pairwise similarity function, named Biological Jaccard Index, that leverages the primary sequence to score the resemblance of a protein pair. In more detail, given two protein $u$ and $v$ and their associated protein sequences $S_u$ and $S_v$, we first identify the set of k-mers~\footnote{K-mers are substrings of length $k$ contained within a general biological sequence.} of each sequence $\hat{S}_u$ and $\hat{S}_v$ and we define the Jaccard Similarity between sets of k-mers as:

\begin{equation}
    \hat{J}(u,v) = \frac{ \mid \hat{S}_u \cap \hat{S}_v \mid }{\mid \hat{S}_u \cup \hat{S}_v \mid}
    \label{eq:biological_jaccard_index}
\end{equation}

To statistically validate if this index was able to keep evolutionary relationships among set of proteins (Gene Duplication Phenomena), we followed the same approach discussed in Section \ref{sub:jac_as_gene_dup}. As shown in supplementary figure \ref{fig:duplicated_gene_set_biological_jaccard_index}, duplicated gene sets have mean jaccard similarity grater than random expectation. Furthermore, the mean of pairwise similarities of each group is greater if Biological Jaccard Index is considered instead of the topological index. This finding might be related to the incompleteness of the interactome.  

\begin{figure}[ht]
\centering
\includegraphics[width=\textwidth]{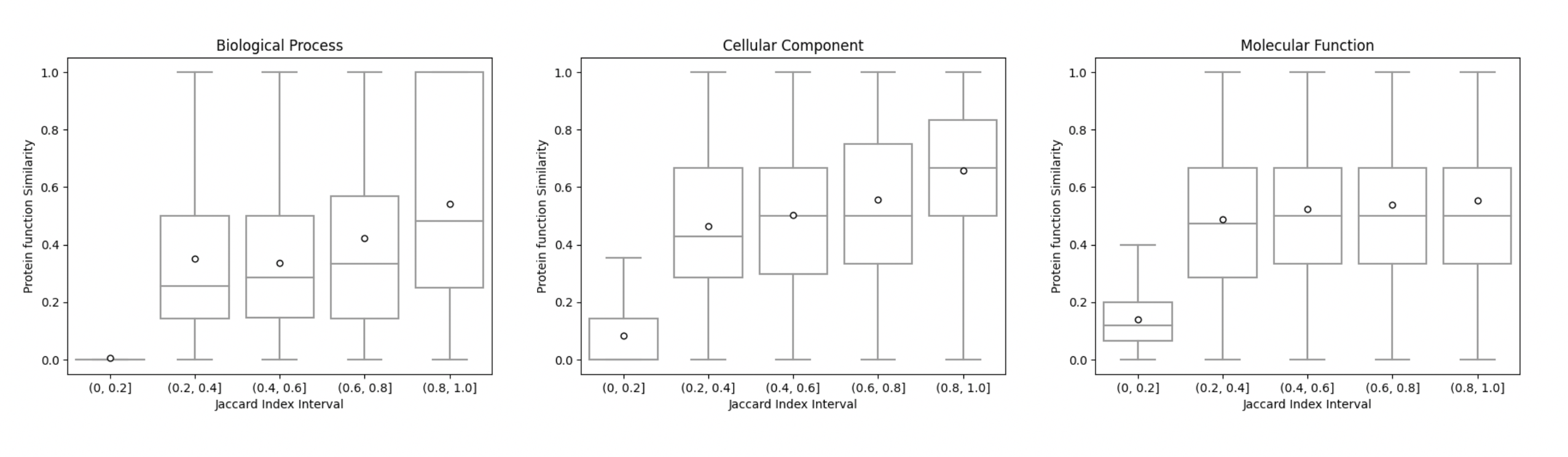}
\caption{Correlation between Sequence Similarity and Functional Similarity. Protein pairs with high Sequence Similarity tend to be involved in the same biological processes and be localized in the same cellular components.}
\label{fig:biological_jaccard_index_functional_similarity}
\end{figure}

Finally, as shown by several studies\cite{hegyi1999relationship,joshi2007quantitative,sangar2007quantitative}, protein pairs with higher sequence similarity tend to be more functionally similar than dissimilar proteins. Indeed, Figure \ref{fig:biological_jaccard_index_functional_similarity} shows the correlation between the Biological Jaccard Index and the functional similarity of a protein pair. We considered Gene Ontologies (Biological Process, Cellular Component and Molecular Function) downloaded from the Gene Ontology Consortium and we plotted the distribution of pairwise Gene Ontology similarity over the interval of the Biological Jaccard Index. All protein pairs with a Jaccard Index in the interval $(0.8, 1.0]$ have several Gene Ontologies in Common if compared with pairs with low JI.

\subsubsection{Model Definition}
\label{sub:model}
From the discussion introduced in sections~\ref{sub:jac_as_protein_interface_sim},~\ref{sub:jac_as_gene_dup} and~\ref{sub:jac_for_seq_sim}, the Similarity between a protein pairs can be used to measure the interface similarity, paralogy and functional similarity. In other words, given a protein pair $(u,v)$, we can use the Jaccard functions to compute the similarity between $v$ and $u$'s neighbors. If $v$ is highly topologically or biologically similar to at least one neighbour of $u$, then there is a good chance that $v$ shares the same binding sites (i.e $v$ has complementary binding site of $u$) or that it is involved in the same function of a neighbour of $u$. 
Thus, we can  model the probability of connection between $u$ and $v$ in two different ways: using the topological Jaccard Index or the Jaccard Index defined in equation \ref{eq:biological_jaccard_index}.

\begin{figure}[h]
\centering
\includegraphics[width=\textwidth]{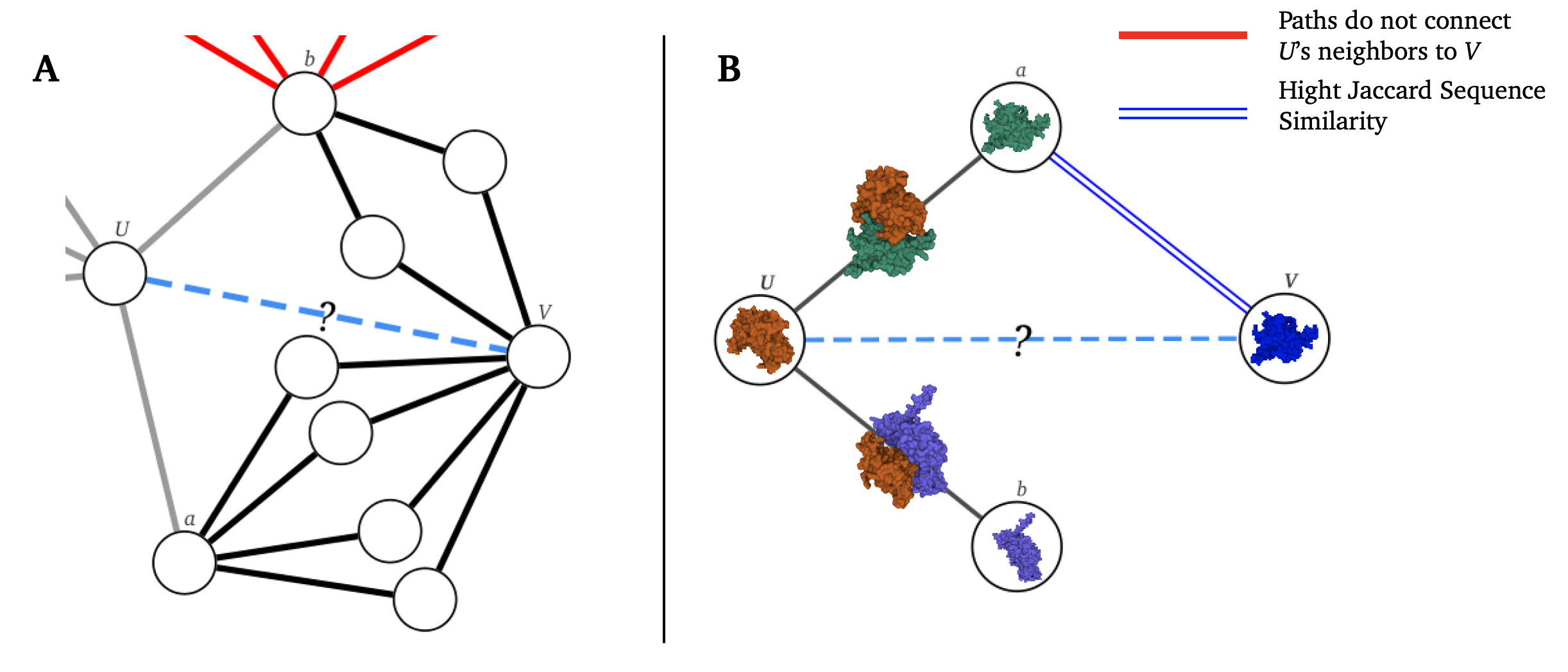}
\caption{Topological and Biological Models of our framework: \textbf{a)} shows a candidate protein pair $(U,V)$ in which $V$ is topologically similar (i.e High Topological Jaccard similarity) to $a$, a (i.e $U$'s neighbour). \textbf{b)} shows a candidate protein pair $(U,V)$ in which $V$ is highly biologically similar to $U$'s neighbour.}
\label{fig:model}
\end{figure}

Figure \ref{fig:model} shows how we score the likelihood of interaction between $u$ and $v$. Figure \ref{fig:model} \textbf{A)} illustrate the model from a topological point of view in which protein $v$ is very similar to protein $a$ (i.e. $u$'s neighbor) and consequently, as discussed previously, $v$ may share complementary binding site with $u$. On the other side, figure \ref{fig:model} \textbf{B)} shows the model from a biological point of view in which, instead of considering the topological Jaccard Similarity, Jaccard Sequence similarity is used to understand if $v$ is functionally similar to at least a neighbor of $u$.

Indeed, as shown in figure \ref{fig:model} \textbf{B)}, $v$ has similar tertiary structure to $a$, but not with $b$. In conclusion, we can formalize the model in the following way:
\begin{equation}
    Max_{sim}(u,v) = max_{a \in N(u)} J(a,v) + max_{b \in N(v)} J(b,u)
\label{max_sim_definition}
\end{equation}

Where $N(i)$ is the set of direct neighbors of i and we can replace the Jaccard Index J, with the one analyzed in section \ref{sub:jac_for_seq_sim}. Surprisingly, the protein pair scoring function that leverages protein's primary structures is not linked to the topology of the PPI network. Indeed, it is able to rank candidate interacting pair that are not close (i.e Shortest path length greater than 3) in the network. Thus, is not affected by the bias induced by network topology and by the network incompleteness (i.e missing data).

\section{Results}
\label{sec:results}

\subsection{Evaluation-Scheme and Accuracy-Measures}
\label{sub:eval}
To evaluate the prediction of our frameworks and comparing it with other algorithms, we selected several Protein-Protein Interaction Networks: 
\begin{itemize}
    
    \item From STRING\cite{szklarczyk2015string} we downloaded PPI networks of different organism (Yeast, C. Elegans, Arabidopsis and Humo Sapiens) and we selected only experimental validated physical interactions. In other words, we removed those protein - protein interactions with an overall score lower than 900.
    
    \item We downloaded a Human Interactome from from \cite{luck2020reference}, that consists of experimental validated PPIs using Yeast to Hybrid screening\cite{}. From BioGRID\cite{}, we downloaded another Human Interactome only considered ”physical” and proteins assigned to the specific to the studied species. Finally, we also included the Interactome3D\cite{} dataset, summarizing currently available interactions with structural evidence

\end{itemize}

Finally, for each protein in each Interactome, we downloaded its associated protein sequence from Uniprot Knowledge Based\cite{}, and in each network, we removed those proteins associated with more than one protein sequence (i.e. those proteins not manually curated by Swiss Prot. Institute\cite{}). The complete list of PPI Networks used in this manuscript is visible in Supp. Table \ref{tab:networks}. To assess accuracy in predicting protein interactions, we performed 10-fold cross validation on each dataset, picking a portion of edges at random to create the Training Graph (90\%) and using the remaining edges (10\%) to Test algorithm's performances. To assess accuracy, we considered four standard prediction indices in Data Mining. For a given set of truly interacting pairs (Test Set),
and an algorithm ranking protein pairs with respect to their interaction likelihood:
\begin{itemize}
    \item \textbf{Precision@500:} this is the fraction of the test set 
    that is successfully retrieved in the top $k$ positions of the ranking computed by the
    algorithm.
    \item \textbf{nDCG\cite{wang2013theoretical}:} The normalized Discounted Cumulative Gain (nDCG) is proven to be able to select the better ranking between any two, substantially different rankings. For binary classification the nDCG is given by:
    \begin{center}
        $\frac{\sum_{i \in P} \frac{1}{log_2(i + 1)}}{\sum_{i = 0}^{|P|} \frac{1}{log_2(i + 1)}}$
    \end{center}
    
Where summation in the numerator runs over all positive instances, while summation in the denominator quantifies the ideal case, in which positive instances appear in the top positions of the algorithm's ranked list.

\end{itemize}

For the seek of completeness, we also compare all the  algorithms with the \textbf{AUROC} and \textbf{AUPRC}, that are widely used to compare algorithm's performances, even if it is well established that \textbf{AUROC} measure overestimates algorithm's performances\cite{}.

\subsection{Competing Methods}
\label{sub:competing_methods}
Network-based approaches leverage network topology to estimate the likelihood of protein interactions. The advantages of network-based methods are high efficiency, easy access to input data (only network topology is needed), and good generalization schemes. 
Table~\ref{tab:methods} collects all PPIs prediction methods that have been implemented and tested.

\begin{table}[htbp]
    \centering
    \begin{tabular}{|p{3.6cm}|p{4cm}|p{6cm}|p{2cm}|}
     \hline
     \multicolumn{4}{|c|}{PPIs Prediction Methods} \\
     \hline
     \textbf{Method}& \textbf{Reference}&\textbf{Index}&\textbf{Path Length}\\
     \hline

     Common Neighbor   & Newman (2001)~\cite{newman2001}    & $CN_{i,j} = |\Gamma_i \cap \Gamma_j|$ &  l = 2\\
     
     Jaccard Index &   Jaccard (1912)~\cite{jaccard1912}  & $JI_{i,j} = \frac{|\Gamma_i \cap \Gamma_j|}{|\Gamma_i \cup \Gamma_j|}$   & l = 2\\
     
     Adamich Adar &Adamic and Adar (2003)~\cite{aa} & $AA_{i,j} = \sum_{z \in \Gamma_i \cap \Gamma_j}\frac{1}{log_2k_z}$ &  l = 2\\
     
     Resource Allocation & Zhou et al. (2009)~\cite{zhou2009predicting}  & $RA_{i,j} = \sum_{z \in \Gamma_i \cap \Gamma_j}\frac{1}{k_z}$   & l = 2\\

     CH1\_L2 & Cannistraci et al. (2018)~\cite{muscoloni2018local}  & $CH1_{L2}(i,j) = \sum_{z \in \Gamma_i \cap \Gamma_j} \frac{ki_z}{k_z}$   & l = 2\\
     
     CH2\_L2 & Cannistraci et al. (2018)~\cite{muscoloni2018local}  & $CH2_{L2}(i,j) = \sum_{z \in \Gamma_i \cap \Gamma_j} \frac{1 + ki_{z}}{1 + ke_z}$   & l = 2\\
     
     %~&~&~&~\\
     
     A3 & Barabâsi et al. (2019)~\cite{kovacs2019network}  & $A3_{i,j} = \sum_{u,v} a_{iu} \cdot a_{u,v} \cdot a_{vj}$   & l = 3\\
     
     L3 & Barabâsi et al. (2019)~\cite{kovacs2019network}  & $L3_{i,j} = \sum_{u,v}\frac{a_{iu} \cdot a_{u,v} \cdot a_{vj}}{k_u \cdot k_v}$   & l = 3\\
     
     L3E & Yuen. (2020)~\cite{yuen2020better}  & $P^{L3E}_{i,j} = \frac{|U|}{|\Gamma_i|} \cdot \frac{|V|}{|\Gamma_j|} \cdot \sum_{U,V} J (i,v) \cdot J(u,j)$   &  l = 3\\
     
     CH1\_L3 & Cannistraci et al. (2018)~\cite{muscoloni2018local}  & $CH1_{L3}(i,j) = \sum_{z_1,z_2 \in L_3} \bigg(  \frac{ki_{z_1} \cdot ki_{z_2} }{k_{z_1} \cdot k_{z_2}} \bigg)^{\frac{1}{2}}$   & l = 3\\
     
     CH2\_L3 & Cannistraci et al. (2018)~\cite{muscoloni2018local}  & $CH2_{L3}(i,j) = \sum_{z_1,z_2 \in L_3} \bigg( \frac{ki_{z_1} \cdot ki_{z_2} }{ke_{z_1} \cdot ke_{z_2}} \bigg)^{\frac{1}{2}}$   & l = 3\\

      SIM & Chen et al. (2020)~\cite{chen2020protein}  & $SIM_{i,j} = \sum_{v \in \Gamma(j)} A \cdot J_{vi} + \sum_{u \in \Gamma(i)} A \cdot J_{uj}$   & l = 3\\
      
      %~&~&~&~\\
      
      Preferencial Attachment &   Barabâsi et al. (2002)~\cite{barabasi2002}  & $PA_{i,j} = k_i \cdot k_j$ & \textit{any}\\
      
      MPS(B) &   Wang et al. (2002)~\cite{wang2021assessment}  & None & \textit{any}\\
      
      SPRINT &   Li, et al. (2020)~\cite{li2017sprint}  & None & \textit{any}\\

     \hline
    \end{tabular}
    \caption{All PPIs prediction methods that have been implemented and tested on several PPI Networks.
    $k$, $ki$~\cite{muscoloni2018local}, and $ke$~\cite{muscoloni2018local} represent the degree, the internal degree, and the external degree of a node in the network.}
\label{tab:methods}
\end{table}

\paragraph{The Preferential Attachment method:} The basic premise is that the probability that a new link has node \textit{i} as an endpoint is proportional to $|\Gamma(i)|$, the current number of neighbors of \textit{i}.

\paragraph{Methods Based on Paths of Length 2:} These heuristics leverage the idea, very popular in social network mining, that two nodes $i$ and $j$ that share many common neighbors are more likely to interact. The most relevant metrics that are based on this hypothesis are: \textbf{Common Neighbor} (CN), defined as the number of common interacting nodes between $i$ and $j$, \textbf{Jaccard Index}\cite{jaccard1912}(JC), a commonly used similarity. Intuitively, it measures the probability that a feature $f$ that \emph{either} $x$ or $y$ possess is actually shared by both. In our case, a feature $f$ is an interacting protein. \textbf{Resource Allocation} \cite{zhou2009predicting}, a similarity index that indicates how well a node $i$ can transmits information to node $j$ using its neighborhood. Finally, Cannistraci et al.\cite{muscoloni2018local} designed a family of metrics ($CH1\_L2$ and $CH2\_L2$), based on \textbf{Resource Allocation}, in which a node $z \in \Gamma_i \cap \Gamma_j$ is associated an internal degree $ki_z$ that is the number of links that common neighbors $i$ and $j$ share between them, and an external degree $ke_z$ that is defined as the number of links that $z$ shares with nodes that are not common neighbors of $i$ and $j$.

\paragraph{Methods Based on Paths of Length 3:} 
These methods are based on the hypothesis that  proteins should have complementary binding sites to be able to interact with each other. Following this idea, Barabasi et al.\cite{kovacs2019network} showed that nodes that share a large number of paths of length $3$ share complementary tertiary structures and are more likely to interact. This idea, i.e., the \textbf{L3 principle}, is at the basis of several heuristiscs proposed since Barabasi et al.'s work. Indeed, Chen et al.\cite{chen2020protein} proposed a network-based link prediction method, named Sim, for PPI networks. This index is designed from two perspectives: the complementarity of protein interaction
interfaces and gene duplication. Cannistraci et al.\cite{muscoloni2018local} designed a family of metrics ($CH1\_L3$ and $CH2\_L3$) that plug in the concept of \textbf{Local Community Paradigm} in the L3 formula designed by Barabasi et al. (2019). Finally, Yuen et al. \cite{yuen2020better} designed a metrics based on the L3 principle and uses Jaccard Index as a penalty to score a candidate interaction.

\paragraph{Methods Based on Protein Primary Structure:}SPRINT is a well know method to predict protein's interactions using their amino-acid sequence. It relies on the idea that proteins similar with interacting proteins are likely to interact as well. In a way or another, this is essentially the idea behind the brute force calculation of PIPE as well as the machine learning algorithms of Martin, Shen, and Guo. Since, all these methods relies on the same concept, we decided to take in consideration SPRINT that has been shown to have a better prediction power and to be faster than the other algorithms.

\paragraph{Methods Implemented for Protein Interaction prediction Challenge:} The proposed Methods has been implemented to participate to the Protein Interaction prediction Challenge organized by the Network Medicine Consortium \footnote{\url{https://www.network-medicine.org/}}. At the challenge, we also proposed a biological score, named MPS(B), to estimate the likelihood of protein interaction that leverages protein sequence in a different way from what we have discussed so far. We took inspiration from PIPE\cite{pitre2006pipe} and we designed a simplified and faster framework in which we don't directly consider all co-occurrences of two sub sequence pairs $(u_i,v_j)$, but we use them to define a protein similarity measure.

\subsection{Algorithm Comparison}
\label{sub:algorithm_comp}
\begin{figure}[t]
\centering
\includegraphics[width=\textwidth]{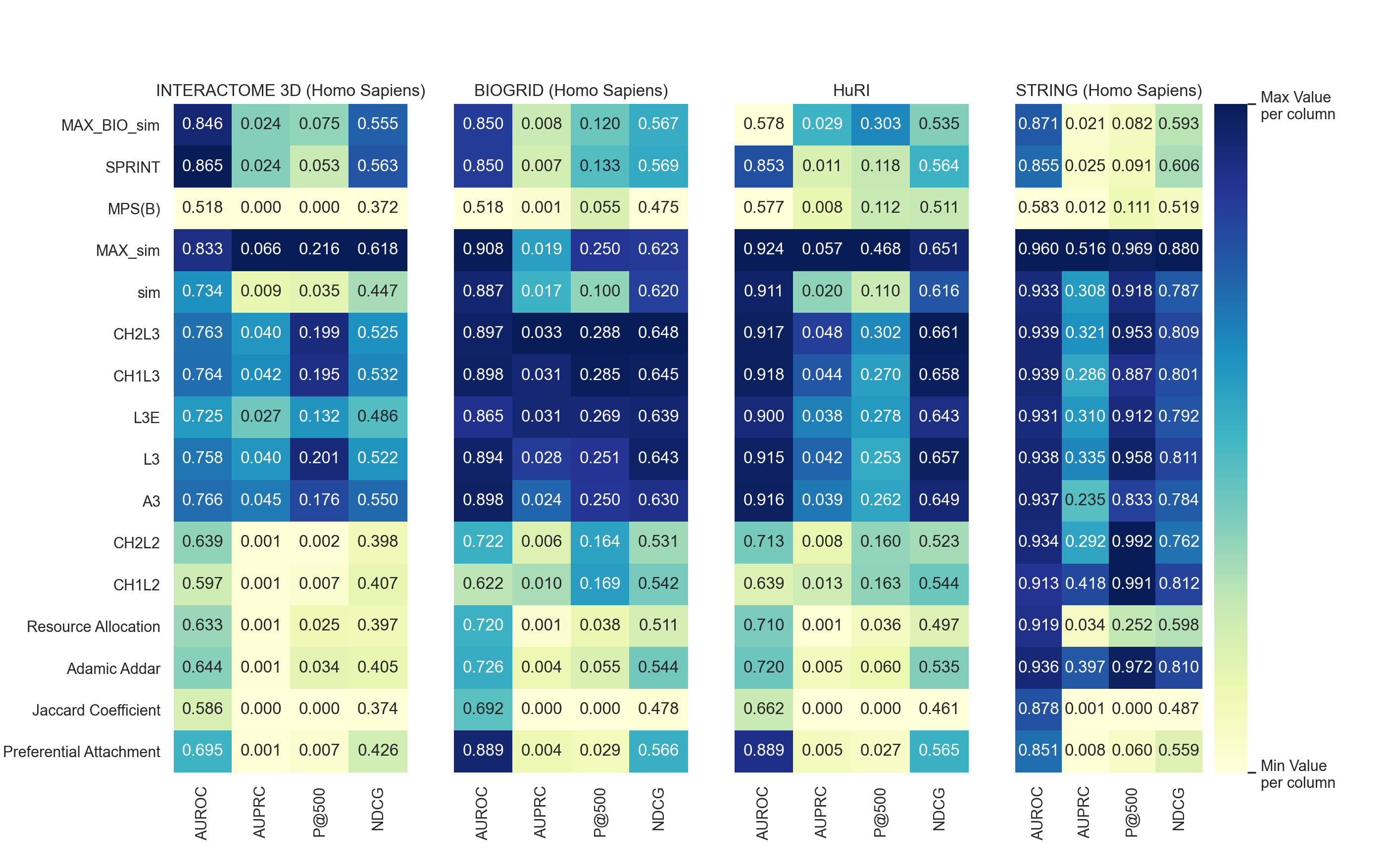}
\caption{ Heatmap plots show the performance of each method on each
Human interactome with the following evaluation metrics: Area Under the Receiver Operating Characteristic (AUROC), Area Under the Precision-Recall Curve (AUPRC), Precision of the top500 predicted PPIs (P@500) and Normalized Discounted Cumulative Gain (NDCG).}
\label{fig:heatmap_comparison}
\end{figure}

This section presents the overall comparison of sixteen different methods using several Human and Not-Human Protein-Protein Interaction networks.

To begin, Figure \ref{fig:heatmap_comparison} shows the algorithm's performances on four different Human PPI networks. If we look at frameworks that leverage only topological information, we can see that algorithms based on the L3 paradigm outperform metrics that can predict only interactions between proteins within two hops (TCP paradigm). Indeed, this pattern is visible on all the data sets considered: $Max^{t}_{sim}$, sim, CH2L3, CH1L3, L3E, L3, and A3 consistently outperform Resource Allocation, Adamich Addar, Jaccard Coefficient, CH1L2, and CH2L2. Furthermore,  $Max^{t}_{sim}$ outperforms all the other heuristics in terms of AUPCR on three networks (INTERACTOME 3D, HuRI, and STRING) and returns the highest number of true positives (Precision@500) on two PPI networks (INTERACTOME 3D, and HuRI). Interestingly, Methods that leverage protein's primary structures show lower computational performances than topology-based frameworks .   

\begin{figure}[h]
\centering
\includegraphics[width=\textwidth]{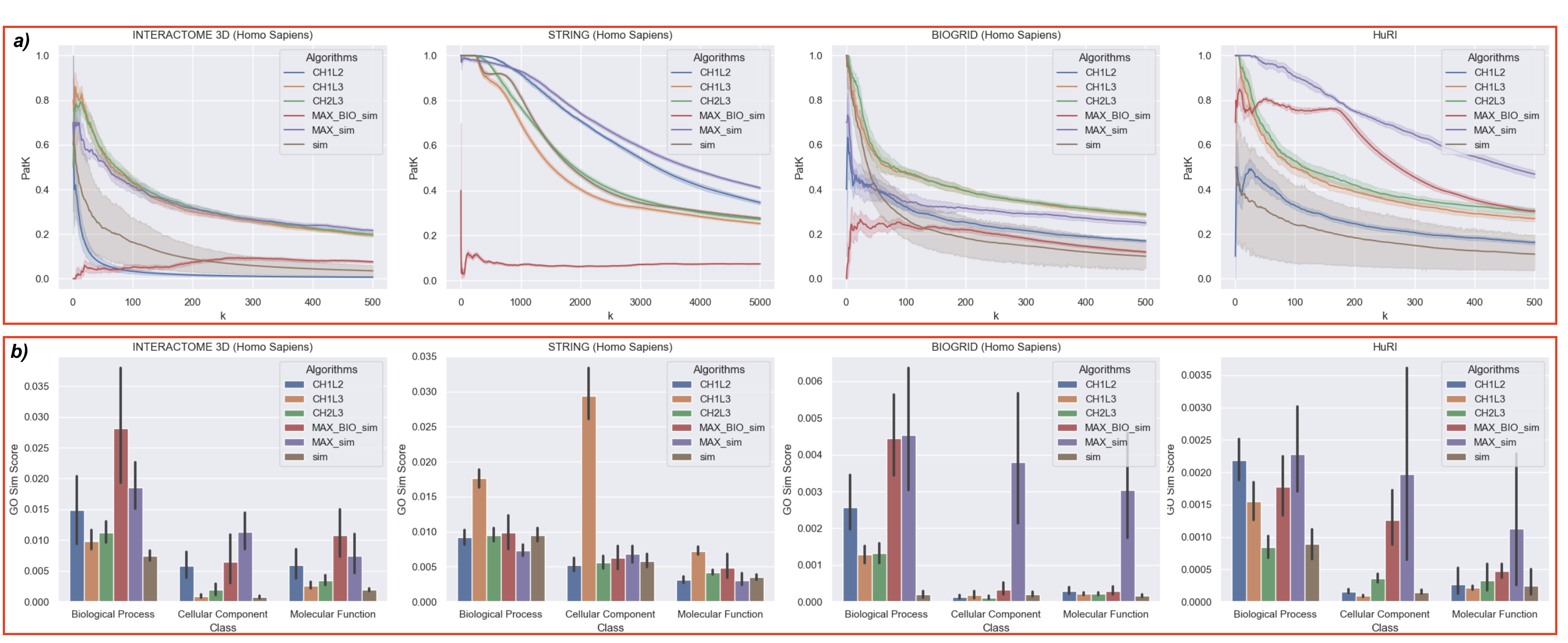}
\caption{Algorithm Comparison. CH1L2, CH1L3, CH2L3, $Max^{b}_{sim}$, $Max^{t}_{sim}$, and sim are are compared using two different measures: \textbf{a)} shows the Precision@k plot and \textbf{b)} shows the MeanGOSimScore. }
\label{fig:algorithm_comparison}
\end{figure}

Since most topology-based algorithms show similar values of P@500, we plotted the precision curve of the best performers. Figure  \ref{fig:algorithm_comparison} \textbf{a)} shows the precision of the best algorithms when the number of top predictions is varying. On the INTERACTOME 3D CH1L3, CH2L3 and $Max^{t}_{sim}$ show similar behavior, but $Max_{sim}$ beats when we consider the top 500 predicted interactions. The same signal is visible on STRING: our topological framework is the best predictor when we consider a larger number of top ranked candidate protein pairs. Finally, on HuRI, $Max^{t}_{sim}$ is the best oracle. Surprisingly, $Max^{b}_{sim}$ shows promising results on this network. This result might be related to the positive correlation between Jaccard Index and Sequence similarity among protein pairs in HuRI as shown in Luck, K. et al.\cite{luck2020reference}.

Besides, we verify the validity of our approaches also on not Human PPI networks. Supp. Figure \ref{fig:heatmap_not_human_ppi} shows the framework's validation on a Synthetic, and three different not Human (Yeast, C. Elegans, and Arabidopsis) interactomes. Results are coherent with the previous ones: L3 approaches outperforms methods that only predict candidate protein pairs within 2 hops(i.e. Resource Allocation, Jaccard Coefficient and Adamich Adar). Furthermore, methods that leverages protein sequence information shows better performances on these networks than Human interactomes. This observation may be due to a greater incompleteness of the Human interactomes than the non-Human ones\cite{hart2006complete}.

To further validate these heuristics, we compared their top candidate protein's interaction using Gene Ontologies annotation\cite{ashburner2000gene,gene2019gene} (Biological Process, Molecular Function, and Cellular Component) downloaded from GO Consortium\cite{gene2015gene}. These annotations are routinely applied to validate computationally predicted links in the literature\cite{you2010using} instead of performing high-throughput validations or pairwise testing experiments. To induce a biological score for each framework, we consider the top 500 candidate pairs of each framework and we compute the GO sim score as in Kovaks et al.\cite{kovacs2019network}. More formally, for each candidate protein pair ($u,v$) in the top k positions we defined $T_u$ and $T_v$ the sets of annotation u and v are respectively involved into, and we estimate their similarity as:
\begin{equation}
    GOSimScore(u,v) = \min_{ i  \in T_u \cap T_v } \frac{2}{n_i}
\end{equation}

Where $n_i$ are the total number of protein involved in annotation $i$. Finally, the framework's score consists of the mean of all $GOSimScore$ associated to the top k candidate predicted protein pairs:

\begin{equation}
    GOSimScore =\frac{1}{k} \cdot \sum_{(u,v) \in Top_k} GOSimScore(u,v)
\end{equation}

Figure  \ref{fig:algorithm_comparison} \textbf{b)} shows the performances of some of the best network-based and sequence-based approaches. Despite their prediction power shown in figure \ref{fig:heatmap_comparison}, candidate pairs seem to be less biologically similar than those ranked by Resource Allocation and Adamich Adar (see Supp. Figure \ref{fig:heatmap_GO_sim_score}). Kovaks et al.\cite{kovacs2019network} observed the same pattern when they analyzed TCP-based methods and compared them with L3 using the same similarity metric. However, It is worth noticing that SPRINT get the best performances if compared with all the other frameworks, and almost all predicted interactions do not share a large number of common neighbors if compared with Resource Allocation (see Supp. Fig. \ref{fig:CN_SPRINT}). In conclusion, if we take in consideration methods based on L3 principle, $Max^{t}_{sim}$ and $Max^{b}_{sim}$ return candidate pairs more similar than those ones returned by other proposed methods (A3, L3, L3E, CH1L3, CH2L3 and sim) as shown in Figure  \ref{fig:algorithm_comparison} \textbf{b)} and Supp. Figure \ref{fig:heatmap_GO_sim_score}.

\subsection{An Unsupervised Approach to Combine Sequence Based and Topological Information}
\label{sub:combiner}

Section \ref{sub:algorithm_comp} allowed us to understand framework's limitations. First,  topology-based approaches are unable to predict candidate protein pairs far away in the Interactome (i.e., Shortest path distance greater than 3). Secondly, as shown in Fig \ref{fig:heatmap_comparison} and Supp. Fig. \ref{fig:heatmap_GO_sim_score}, they are very biased on the topology of the Protein-Protein Interaction network and fail to rank highly similar candidate protein pairs. On the other hand, methods that rely on primary sequence can score the likelihood of interaction of candidate proteins localized in distant areas of the network. Also, they rely on the information of the protein's primary structure, reducing the bias of the Interactome.

Consequently, we investigate the effect of combining topological and biological frameworks to predict new protein interactions to moderate their topological bias and rank biologically similar protein pairs.  Given the topological feature $f_t$ and the biological feature $f_b$, we combine them using a linear combination of the two:
\begin{equation}
     f_{t,b}(u,v) = \alpha \cdot f_t + \beta \cdot f_b
\end{equation}
To choose the value of  $\alpha$ and $\beta$, we first normalized/standardized $f_t$ and  $f_b$. Then, we projected them on a 1-dimensional space with the help of principal components analysis (PCA), a classical technique for extracting patterns and performing dimensionality reduction from unlabeled data. It computes a linear combination of two features, forming the direction that captures the most variance in the data set.

\begin{figure}[h]
\centering
\includegraphics[width=\textwidth]{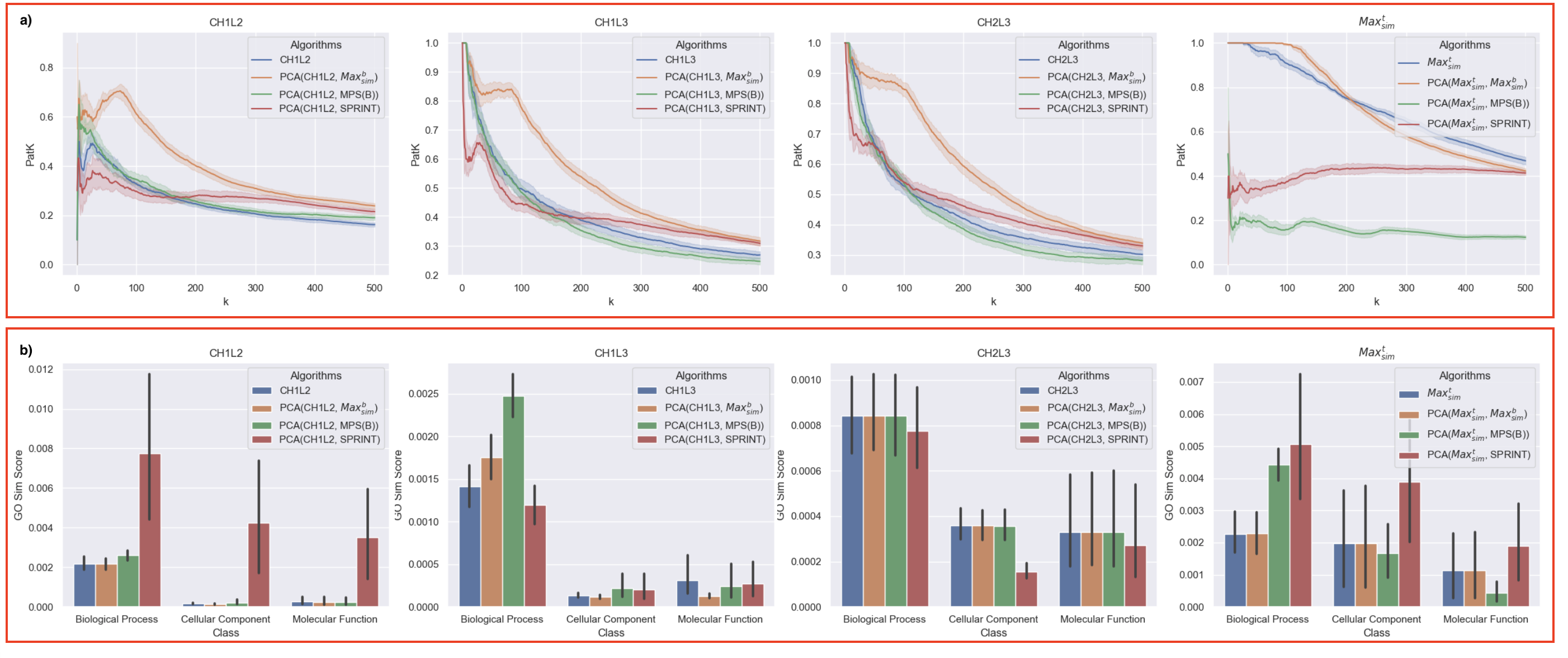}
\caption{Algorithm Comparison. CH1L2, CH1L3, CH2L3, $Max^{b}_{sim}$, $Max^{t}_{sim}$, and sim are are compared using two different measures: \textbf{a)} shows the Precision@k plot and \textbf{b)} shows the GOSimScore. }
\label{fig:combination_performer}
\end{figure}

Figure \ref{fig:combination_performer} illustrates the performance of the combined metrics compared to individual ones. We analyzed the performances on HuRI Interactome: The network in which both topological and biological metrics have an excellent prediction power on P@K and GoSimScore. 
Each plot represents a topological framework and its combinations. 
As a first result, $Max^{b}_{sim}$ can improve the retrieval of the top 200 candidate protein pairs of each topological algorithm considered, as shown in Figure \ref{fig:combination_performer} \textbf{a)}. Furthermore, biological frameworks help network-based approaches to retrieve more biologically similar protein pairs, as shown in Figure \ref{fig:combination_performer} \textbf{b)}.

\section{Conclusion}
\label{sec-conclusions}

We presented two network-based approaches, named $Max^{t}_{sim}$ and $Max^{b}_{sim}$, for predicting candidate interacting protein pairs. The exceptional success of our models depends on their ability to capture the structural and evolutionary principles that drive protein-protein interactions that Jaccard Indices infer, as discussed in Material and Methods. 

We compared $Max^{t}_{sim}$ and $Max^{b}_{sim}$ with state-of-the-art network-based approaches using a 10-fold cross-validation approach on several Human and Not Human Protein-Protein Interaction networks. The ability of network-based approaches to predict protein interactions strongly depends on the network's topology. Indeed, L3 approaches have similar performances, and $Max^{t}_{sim}$ outperforms the other heuristics in the majority of the network considered. On the other hand, Sequence-based approaches that do not rely exclusively on network topology fail in predicting candidate interacting pairs. Furthermore, to better understand the biological similarity of the top candidate protein pairs ranked by each framework, we implemented the GOSimScore as Kovaks et al.\cite{}. Surprisingly, SPRINT, a well-known sequence-based approach, returned the candidate protein pairs involved in similar processes, molecular function, and in the same area.

Surprised by the excellent prediction power of $Max^{b}_{sim}$ on HuRI Interactome, we investigated possible combinations with the best network-based methods. We discover that combining topological and biological methods help in ranking interacting protein pairs without decreasing their biological similarity. 

However, the $Max^{t}_{sim}$ framework is not without limitations. Like all L3-based methods, $Max^{t}_{sim}$ alone cannot find interacting partners for proteins without known links. For such proteins, we integrated information on sequence combining $Max^{t}_{sim}$ with $Max^{b}_{sim}$, SPRINT or MPS(B).

Another limitation concerns the computational validation: Fig. \ref{fig:heatmap_comparison} and Supp. Fig. \ref{fig:heatmap_GO_sim_score}  represent two different ways to validate candidate protein pairs predicted by the different frameworks. It is worth noticing that frameworks based on network topology perform better when compared with classical data-mining measures such as P@K and NDCG. At the same time, frameworks that rely on protein's primary sequence outperform the others when GOSimScore is considered.

We can still do much work to improve our framework: we can think of a more reasonable way to combine topology and sequence-based methods. We can extend the framework to use several biological information such as Co-Expression of protein pairs, their functional similarities, and phylogenetic profile similarity, evolutionary history or 3D structure that we have not considered in this manuscript but that are integrated in several well-known bioinformatics tools\cite{zhang2012structure,keskin2016predicting,szilagyi2005prediction,wuchty2006topology}. In conclusion, $Max^{t}_{sim}$ and $Max^{b}_{sim}$ with their combinations are promising tools for the completion of the human interactome, allowing us to exploit network effects as we aim to uncover the mechanistic roots of human disease\cite{menche2015uncovering, huttlin2017architecture}.

\section*{Acknowledgement}
This work is partially supported by the ERC Advanced Grant 788893 AMDROMA ``Algorithmic and Mechanism Design Research in Online Markets'', the EC H2020RIA project ``SoBigData++'' (871042), and the MIUR PRIN project ALGADIMAR ``Algorithms, Games, and Digital Markets''.

\bibliography{main}

\appendix
\newpage
\section{Appendix}

\subsection{Connection Probability}

\begin{figure}[ht]
\centering
\includegraphics[width=\textwidth]{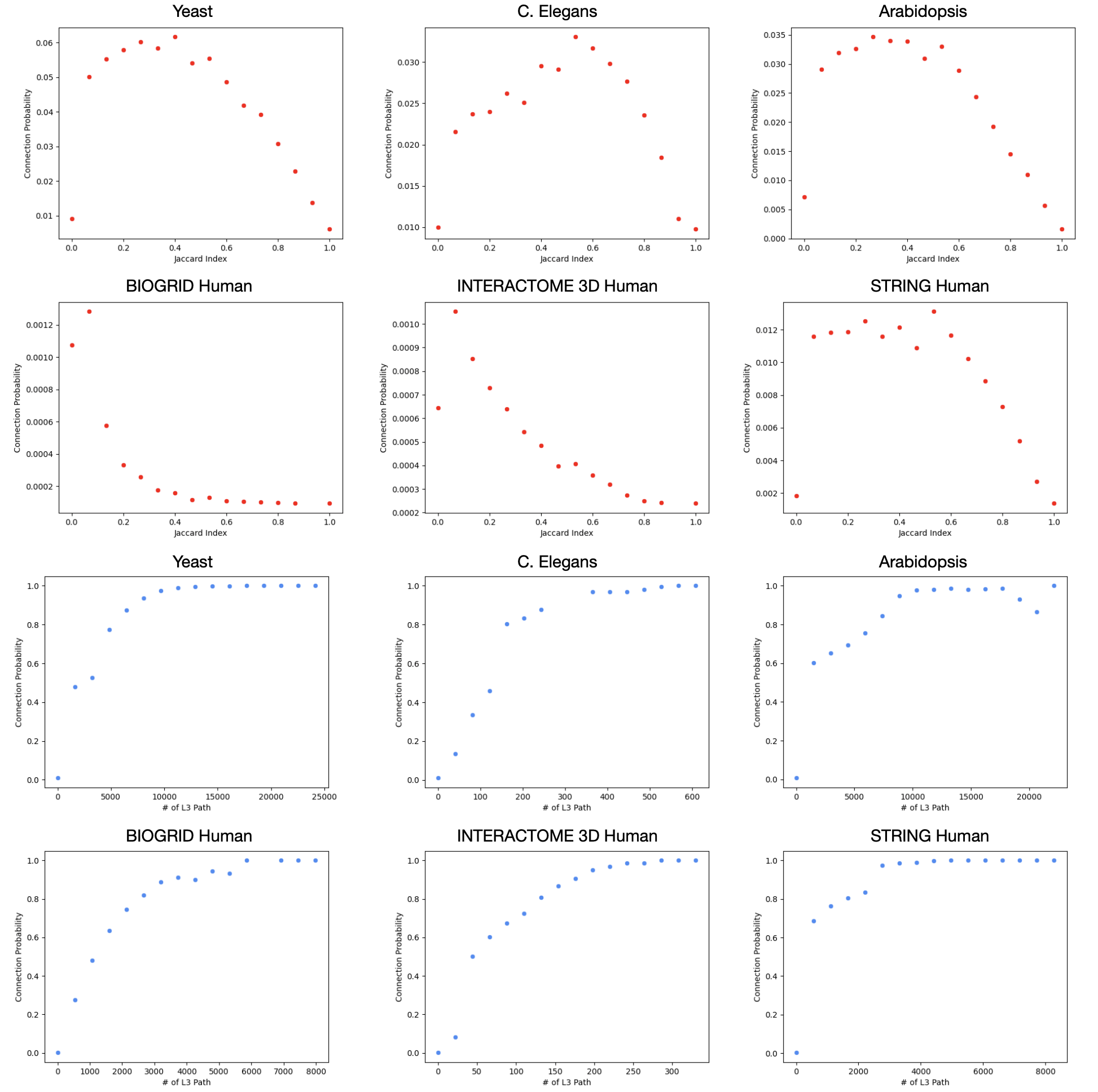}
\caption{Network similarity does not imply connectivity. } 

\label{fig:connection_probability_all_network}
\end{figure}

\newpage

\subsection{Protein-Protein Interaction Network Dataset}

\begin{table}[htbp]
    \centering
    \begin{tabular}{ |p{3cm}|p{3cm}|p{4cm}|p{3cm}|p{3cm}|}
     \hline
     \multicolumn{5}{|c|}{Considered Protein-Protein Interaction Networks} \\
     \hline
     \textbf{Data Set}&
     \textbf{Organism}& 
     \textbf{Reference}&
     
     \textbf{Number of Nodes ($|V|$)}&
     \textbf{Number of Edges ($|E|$)}\\
     \hline
     
        & Synthetic & V{\'a}zquez et al. (2003)\cite{vazquez2003modeling} & $8272$ &  $52922$\\
     
     STRING   &Yeast& Szklarczyk et al. (2015)\cite{szklarczyk2015string} & $2539$ &  $29219$\\
     
     STRING   &C. Elegans& Szklarczyk et al. (2015)\cite{szklarczyk2015string} & $517$ &  $1329$\\
     
     STRING   &Arabidopsis& Szklarczyk et al. (2015)\cite{szklarczyk2015string} & $2831$ &  $28628$\\
     
     STRING   &Homo Sapiens& Szklarczyk et al. (2015)\cite{szklarczyk2015string} & $6325$ &  $36674$\\
     
     BioGRID   &Homo Sapiens& Oughtred et al. (2019)  \cite{oughtred2019biogrid}& $13103$ &  $92250$\\
     
     INTERACTOME 3D   &Homo Sapiens& Mosca et al (2013)\cite{mosca2013interactome3d} & $6100$ &  $11996$\\
     
     HuRI   & Homo Sapiens & Luck, K.et al (2020)\cite{luck2020reference} & $8272$ &  $52548$\\
     
     \hline
    
     \end{tabular}
     \caption{Considered Protein Protein Interaction Networks: A synthetic network created using the method discussed in\cite{vazquez2003modeling} and several networks downloaded from STRING DB\cite{szklarczyk2015string} }
     \label{tab:networks}
\end{table}

\subsection{ Jaccard Index to Model Evolutionary and Functional Similarity}

To statistically quantify if proteins created by this evolutionary process show this behaviour, we downloaded from \cite{ouedraogo2012duplicated} a data set consisting of groups of protein products generated by the gene duplication process. Firstly, we filtered out the smallest groups, keeping only groups consisting of  a large number of proteins (i.e number of proteins $\ge 10$ ) that appear in the PPI network. Secondly, for each group $I$, we defined the Mean Jaccard Index of group $I$ ($MJI_I$) as:
\begin{equation}
    MJI_I = \frac{1}{\mid m \mid} \cdot \sum_{(u,v) \in IxI} \hat{J}(u,v)
    \label{eq:mean_jaccard}
\end{equation}

Where $m$ is the size of $IxI$ and $\hat{J}(u,v)$ is the Biological Jaccard index defined in section \ref{sub:jac_for_seq_sim} of the main article. Finally, we compared the value of each group's Mean Jaccard Index with a random distribution of the score computed using random set of proteins with the same size of the original group $I$.

\begin{figure}[ht]
\centering
\includegraphics[width=\textwidth]{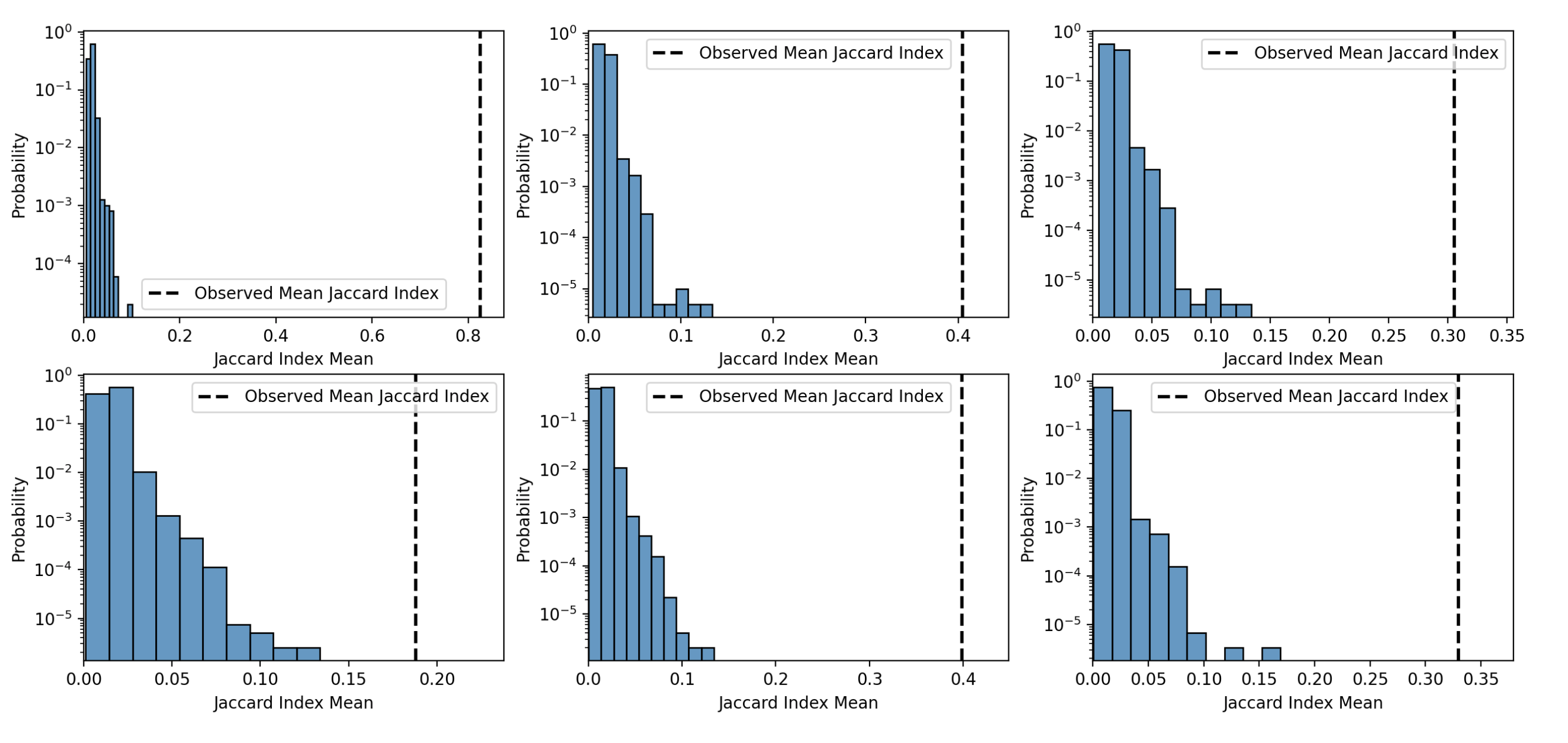}
\caption{Proteins originated by Gene Duplication Phenomena have more similar primary structures than Random Expectation} 

\label{fig:duplicated_gene_set_biological_jaccard_index}
\end{figure}

\subsection{Algorithm Comparison}

\begin{figure}[!htb]
\centering
\includegraphics[width=\textwidth]{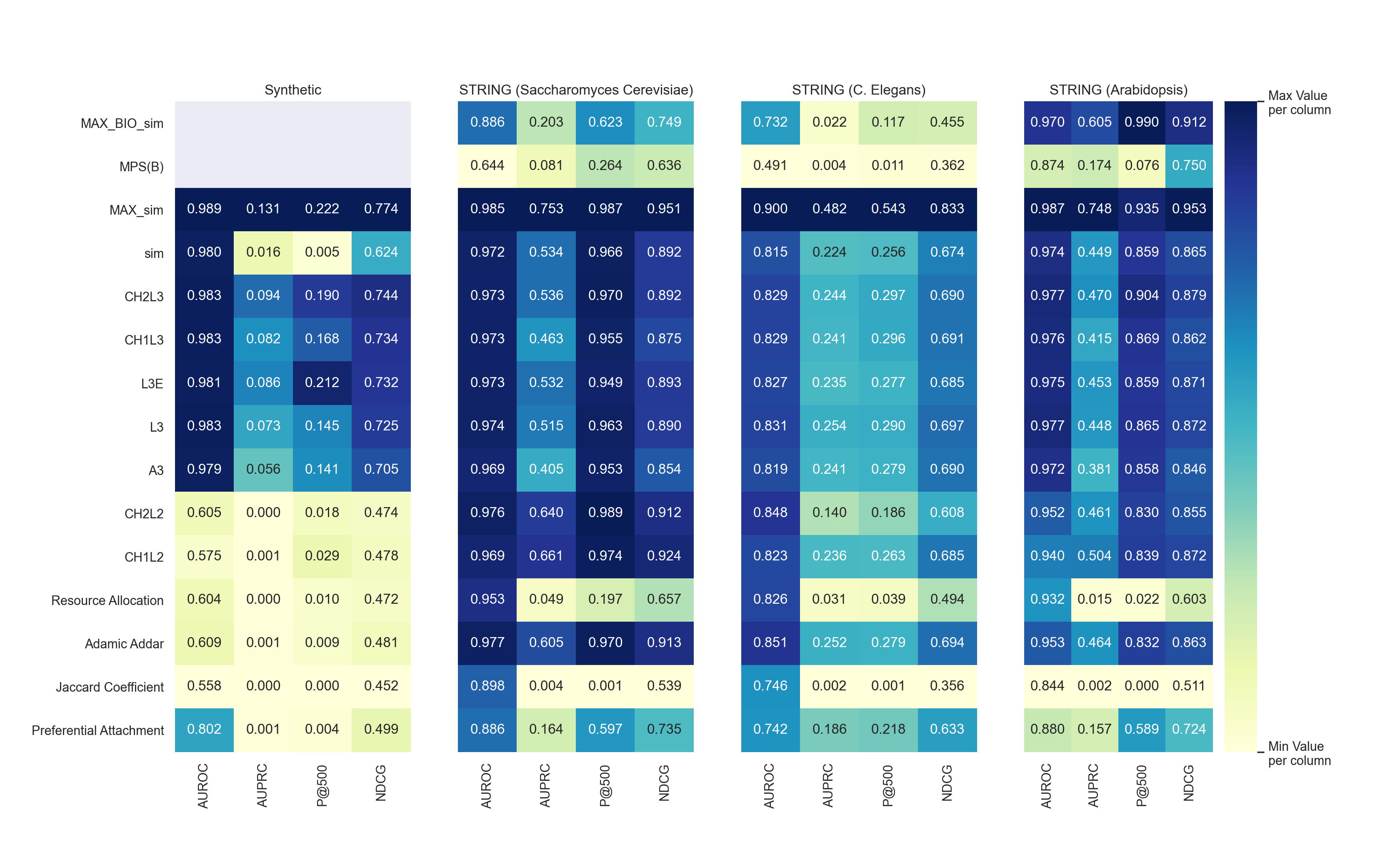}
\caption{Heatmap plots show the performance of each method on each
Human interactome with the following evaluation metrics: Area Under the Receiver Operating Characteristic (AUROC), Area Under the Precision-Recall Curve (AUPRC), Precision of the top500 predicted PPIs (P@500) and Normalized Discounted Cumulative Gain (NDCG).}
\label{fig:heatmap_not_human_ppi}
\end{figure}

\begin{figure}[!htb]
\centering
\includegraphics[width=\textwidth]{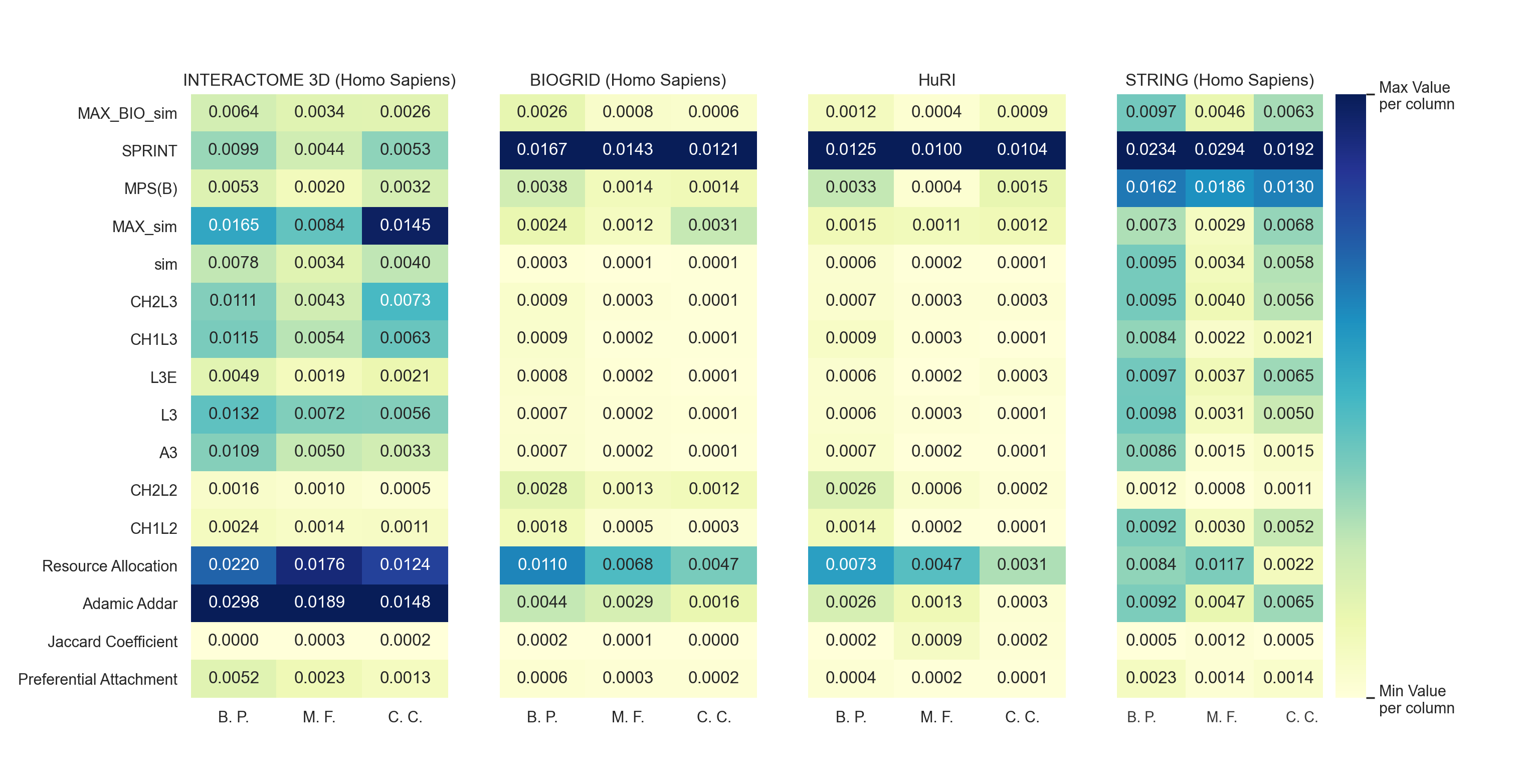}
\caption{Heatmap plots show the performance of each method on each
Human interactome with the GO Sim Score computed on Biological Process (B.P), Molecular Function(M.F) and Cellular Component(C.C) .}
\label{fig:heatmap_GO_sim_score}
\end{figure}

\begin{figure}[!htb]
\centering
\includegraphics[width=\textwidth]{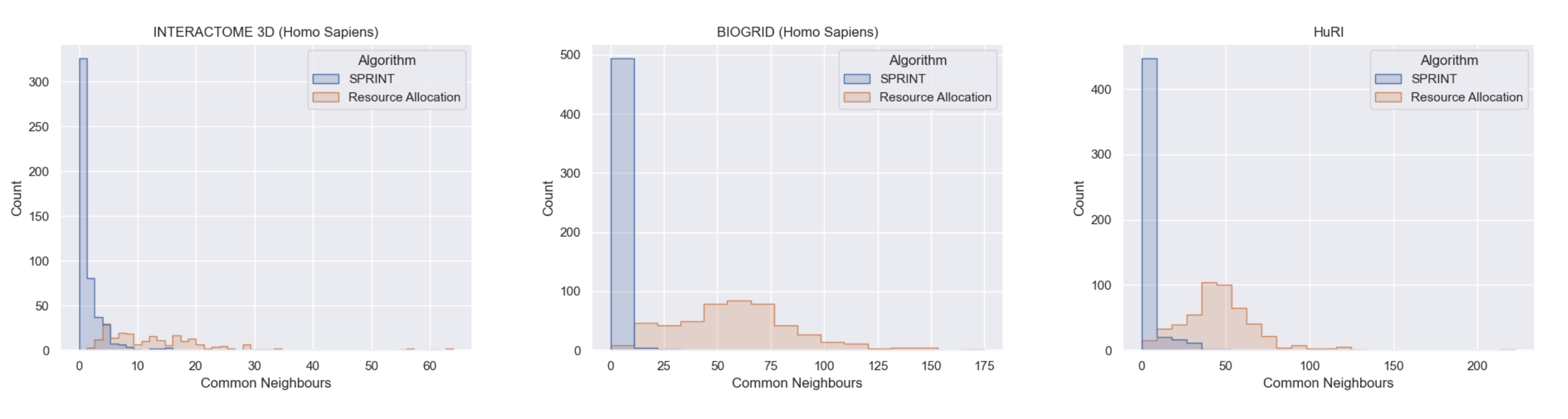}
\caption{Analysis of top 500 candidate pairs predicted respectively by SPRINT and Resource Allocation. For each algorithm we plot the histogram representing the distribution of the number of common neighbours (CN) for the top 500 candidate pairs. It is easy to see that candidate pairs ranked by resource allocation show a greter number of CN if compared with those ones ranked by SPRINT. We have taken in consideration interactomes in which Resource Allocation is one of the best predictors.}
\label{fig:CN_SPRINT}
\end{figure}

\end{document}